\documentclass{article}
\usepackage{times}
\usepackage{amsfonts}
\usepackage{graphicx}
\usepackage[pdfmark]{hyperref}
\begin{document}
\noindent
{\Large A HOLOGRAPHIC MAP OF ACTION ONTO ENTROPY}
\vskip1cm
\noindent
{\bf D. Acosta$^{1,a}$, P. Fern\'andez de C\'ordoba$^{2,b}$, J.M. Isidro$^{2,c}$ and J.L.G. Santander$^{3,d}$}\\
${}^{1}$Departamento de Matem\'aticas, Universidad de Pinar del R\'{\i}o,\\ Pinar del R\'{\i}o, Cuba\\
${}^{2}$Instituto Universitario de Matem\'atica Pura y Aplicada,\\ Universidad Polit\'ecnica de Valencia, Valencia 46022, Spain\\
${}^{3}$C\'atedra Energesis de Tecnolog\'{\i}a Interdisciplinar, Universidad Cat\'olica de Valencia,\\ C/ Guillem de Castro 94, Valencia 46003, Spain\\
${}^{a}${\tt dago@mat.upr.edu.cu}, ${}^{b}${\tt pfernandez@mat.upv.es}\\
${}^{c}${\tt joissan@mat.upv.es}, ${}^{d}${\tt martinez.gonzalez@ucv.es} \\
%\vskip.5cm
%\noindent
%\today
\vskip.5cm
\noindent
{\bf Abstract} We propose a holographic correspondence between the action integral $I$ describing the mechanics of a finite number of degrees of freedom in the bulk, and the entropy $S$ of the boundary (a holographic screen) enclosing that same volume. The action integral must be measured in units of ($i$ times) Planck's constant, while the entropy must be measured in units of Boltzmann's constant. In this way we are led to an intriguing relation between the second law of thermodynamics and the uncertainty principle of quantum mechanics.

%\tableofcontents

\section{Introduction}\label{intro}

There are compelling reasons to believe that quantum mechanics must be an emergent phenomenon \cite{ADLER, CARROLL2, CARROLL3, ELZE1, ELZE2, GROESSING0, GROESSING1, GROESSING2, GROESSING5, THOOFT1, THOOFT2}. Actually not just quantum mechanics, but also gravity and spacetime appear to be emergent phenomenona as well (for a comprehensive review see \cite{PADDY} and refs. therein). The guiding principle in all emergent theories is the fact that they provide a coarse--grained description of some underlying theory \cite{CARROLL1}. Due to our ignorance of a full microscopic description, emergent phenomena are in principle  amenable to a thermodynamical description.

It is the purpose of this contribution\footnote{Talk presented by J.M.I. at the {\it 5th International Heinz von Foerster Congress: Emergent Quantum Mechanics}\/, Vienna, Nov. 11-13,  2011.} to develop an approach to emergent quantum mechanics from the {\it entropic}\/ point of view presented in ref. \cite{ISIDRO}.  We take the view that, apart from other important reasons \cite{THEO, MATONE, KAUFFMAN, KHRENNIKOV, MUNK1, MUNK2, CETTO}, quantum theory must be an emergent phenomenon {\it also}\/ because the spacetime it is defined on is an emergent concept. There exist in the literature a number of different approaches to account for the emergent nature of spacetime, too numerous to quote here in detail. Here we will follow the holographic \cite{HOLO, GRAMA} proposal presented in ref. \cite{VERLINDE}. Thus gravity and quantum mechanics share the common feature of being effective, thermodynamical descriptions of their respective underlying theories.

We should point out that quantum mechanics can be recast in thermodynamical terms \cite{OLAH}, although without making use of the properties of emergence and holography used here. On the contrary, our approach hinges crucially on the notions of emergence and holography. Altogether, our approach will provide us with {\it a holographic, entropic picture of emergent quantum mechanics}\/.

\section{The correspondence}

\subsection{The main result}\label{resultat}

Our main result can be summarised in the holographic correspondence
\begin{equation}
{\rm (bulk)}\;\frac{{\rm i}I}{\hbar}\leftrightarrow\frac{S}{k_B}\;{\rm (boundary)}.
\label{korrespondenzia}
\end{equation}
This correspondence can be explained as follows. Let a finite 3--dimensional volume ${\cal V}$ be given, such that it is bounded by a closed 2--dimensional surface ${\cal S}$ (a holographic screen, see \cite{VERLINDE}). Let a finite number of quantum--mechanical degrees of freedom be defined within ${\cal V}$, described by the action integral $I$. The screen ${\cal S}$ is assumed to carry $N$ information bits. These bits encode the holographic projection, onto ${\cal S}$, of the degrees of freedom within ${\cal V}$. Since we do not know the specific mechanism whereby the holographic principle projects the mechanics within ${\cal V}$ onto its boundary ${\cal S}=\partial {\cal V}$, we assign the screen an entropy $S$, which measures our ignorance about the specific nature of the degrees of freedom on the surface.
Thus $I$ describes a mechanical system in the bulk, while $S$ describes its corresponding thermodynamics on the boundary. In this setup, space merely plays the role of a {\it storage device for  information}\/; space has already emerged within ${\cal S}$, while it does not yet exist outside ${\cal S}$ \cite{VERLINDE}. It will be observed that each side of the dimensionless correspondence (\ref{korrespondenzia}) is measured in units of the corresponding quantum---the quantum of action (Planck's constant $\hbar$) on the mechanical side, the quantum of entropy (Boltzmann's constant $k_B$) on the thermodynamical side. Finally, there is a relative factor of $i$, whose origin will be elucidated presently. For the moment we note that the semiclassical limit  $\hbar\to 0$ in the bulk corresponds to letting $k_B\to 0$ on the boundary. Last but not least, the two quantities $I$ and $S$ separately obey a corresponding extremum principle. Eqn. (\ref{korrespondenzia}) differs from an analogous correspondence, presented in \cite{ISIDRO}, by a factor of 2, to be explained later.

\subsection{A quantum of entropy}\label{aqoe}

The starting point in ref. \cite{VERLINDE} is a classical point particle of mass $M$ approaching a holographic screen ${\cal S}$, from that side of the latter on which spacetime has already emerged. At a distance from ${\cal S}$ equal to 1 Compton length, the particle causes the entropy $S$ of the screen to increase by the amount
\begin{equation}
\Delta S=2\pi k_B,
\label{entropia}
\end{equation}
where $k_B$ is Boltzmann's constant. The above can also be understood as meaning that $2\pi k_B$ is the {\it quantum}\/ by which the entropy of the screen increases, whenever a particle crosses ${\cal S}$. The factor $2\pi$ on the right--hand side is conventional. Relevant is only the fact that the entropy increase of the screen appears quantised in units of $k_B$.

\subsection{Two thermodynamical languages}\label{ttre}

Thermodynamics can be conveniently expressed in either of two equivalent languages, respectively called the energy representation and the entropy representation \cite{CALLEN}. Any given thermodynamical system can be completely described if one knows its {\it fundamental equation}\/. The latter contains all the thermodynamical information one can obtain about the system. The fundamental equation can be expressed in either of two equivalent ways, respectively called the {\it energy representation}\/ and the {\it entropy representation}\/. In the energy representation one has a fundamental equation $E=E(S, \ldots)$, where the energy $E$ is a function of the entropy $S$, plus of whatever additional variables may be required. In the entropy representation one solves for the entropy in terms of the energy to obtain a fundamental equation $S=S(E, \ldots)$.

Here we will argue that quantum mechanics as we know it ({\it i.e.}, on spacetime) corresponds to the energy representation, while quantum mechanics on a holographic screen ({\it i.e.}, in the absence of spacetime) will correspond to the entropy representation.  Our goal is to describe the laws of {\it entropic quantum mechanics}\/, that is, the thermodynamical laws on the boundary ${\cal S}$ that correspond to the quantum mechanics within ${\cal S}=\partial{\cal V}$.

One must bear in mind, however, that standard thermodynamical systems admit both representations (energy and entropy) simultaneously, which representation one uses being just a matter of choice. In our case this choice is dictated, for each fixed observer, by that side of the screen on which the observer wants to study quantum mechanics. For example there is no energy variable beyond the screen, as there is no time variable, but an observer can assign the screen an entropy, measuring the observer's ignorance of what happens beyond the screen. This notwithstanding, the analogy with thermodynamical systems we have just sketched can be quite useful.

\subsection{A (classical) holographic dictionary}\label{aholdic}

Assume that we are give a foliation of 3--space by 2--dimensional holographic screens ${\cal S}_j$: $\mathbb{R}^3=\cup_{j\in {\cal J}}{\cal S}_j$, where the index $j$ runs over some (continuous) set ${\cal J}$. For reasons to be explained presently we will restrict our attention to potentials such that the ${\cal S}_j$ are all closed surfaces; we denote the finite volume they enclose by ${\cal V}_j$, so $\partial {\cal V}_j={\cal S}_j$.

One can formulate a {\it holographic dictionary}\/ between gravitation, on the one hand, and thermodynamics, on the other \cite{VERLINDE}. Let $V_G$ denote the gravitational potential created by a total mass $M=\int_{\cal V}{\rm d}^3V\,\rho_M$ within the volume ${\cal V}$. Then the following two statements are equivalent:\\
{\it i)} there exists a gravitational potential $V_G$ satisfying Poisson's equation $\nabla^2 V_G=4\pi G\rho_M$, such that a test mass $m$ in the background field created by the mass distribution $\rho_M$ experiences a force ${\bf F}=-m\nabla V_G$;\\
{\it ii)} given a foliation of 3--space by holographic screens, $\mathbb{R}^3=\cup_{j\in {\cal J}}{\cal S}_j$, there are two scalar quantities, called entropy $S$ and temperature $T$, such that the force acting on a test mass $m$ is given by $F\delta x=\int_{\cal S}T\delta {\rm d}S$. The latter integral is taken over a screen that does not enclose $m$.\\
Moreover, the thermodynamical equivalent of the gravitational theory includes the following {\it dictionary entries}\/ \cite{VERLINDE}:
\begin{equation}
\frac{1}{k_B}S(x)=\frac{-1}{4\hbar c L_P^2}V_G(x)A(V_G(x)),
\label{sabinebiene}
\end{equation}
\begin{equation}
2\pi k_BT(x)=\frac{{\rm d}V_G}{{\rm d}n},
\label{tivi}
\end{equation}
\begin{equation}
\frac{k_B}{2}\int_{\cal S}{\rm d}^2a\, T=L_P^2Mc^2.
\label{defitermo}
\end{equation}
In (\ref{sabinebiene}), (\ref{tivi}) and (\ref{defitermo}) we have placed all thermodynamical quantities on the left, while their mechanical analogues are on the right. As in ref. \cite{VERLINDE}, the area element ${\rm d}^2a$ on ${\cal S}$ is related to the infinitesimal number of bits ${\rm d}N$ on it through ${\rm d}^2a=L_P^{2}{\rm d}N$. We denote the area of the equipotential surface passing through the point $x$ by $A(V_G(x))$, while ${\rm d}V_G/{\rm d} n$ denotes the derivative of $V_G$ along the normal direction to the same equipotential. The above expressions tell us how, given a gravitational potential $V_G(x)$ and its normal derivative ${\rm d}V_G/{\rm d} n$, the entropy $S$ and the temperature $T$ can be defined {\it as functions of space}\/.

Specifically, eqn. (\ref{sabinebiene}) expresses the proportionality between the area $A$ of the screen ${\cal S}$ and the entropy $S$ it contains. This porportionality implies that gravitational equipotential surfaces get translated, by the holographic dictionary, as {\it isoentropic surfaces}\/, above called holographic screens ${\cal S}$.

Equation (\ref{tivi}) expresses the Unruh effect: an accelerated observer experiences the vacuum of an inertial observer as a thermal bath at a temperature $T$ that is proportional to the observer's acceleration ${\rm d}V_G/{\rm d}n$.

{}Finally, eqn. (\ref{defitermo}) expresses the first law of thermodynamics and the equipartition theorem. The right--hand side of (\ref{defitermo}) equals the total rest energy of the mass enclosed by the volume ${\cal V}$, while the left--hand side expresses the same energy content as spread over the bits of the screen ${\cal S}=\partial {\cal V}$, each one of them carrying an energy $k_BT/2$. It is worthwhile noting that equipartition need not be postulated. Starting from (\ref{tivi}) one can in fact prove the following form of the equipartition theorem \cite{ISIDRO}:
\begin{equation}
\frac{k_B}{2}\int_{\cal S}{\rm d}^2a\, T=\frac{A({\cal S})}{4\pi}U({\cal S}), \qquad A({\cal S})=\int_{\cal S}{\rm d}^2a.
\label{defitermonova}
\end{equation}
Above, $U$ can be an arbitrary potential energy, subject only to the requirement that its equipotential surfaces are closed. We will henceforth mean eqn. (\ref{defitermonova}) when referring to the first law and the equipartition theorem. In all the above we are treating the area as a continuous variable, but in fact it is quantised \cite{VERLINDE}. If $N({\cal S})$ denotes the number of bits of the screen ${\cal S}$, then
\begin{equation}
A({\cal S})=N({\cal S})L_P^2.
\label{numm}
\end{equation}
However, in the limit $N\to\infty$, when $\Delta N/N<<1$, this approximation of the area by a continuous variable is accurate enough. We will see later on that letting $N\to\infty$ is equivalent to the semiclassical limit in quantum mechanics.

\section{The emergence of quantum mechanics}

We intend to write a holographic dictionary between quantum mechanics, on the one hand, and thermodynamics, on the other. This implies that we will need to generalise eqns.  (\ref{sabinebiene}), (\ref{tivi}) and (\ref{defitermonova}) so as to adapt them to our quantum--mechanical setup.  Thus we will replace the classical particle of \cite{VERLINDE} with a quantum particle, subject to some potential energy $U$ of nongravitational origin, but we will take (\ref{entropia}) to hold for a quantum particle as well. We will assume $U$ to be such that its equipotential surfaces ${\cal S}_j$ are closed, in agreement with our assumptions about the foliation. Let $H=K+U$ be the classical Hamiltonian function on $\mathbb{R}^3$ whose quantisation leads to the quantum Hamiltonian operator $\hat H=\hat K+\hat U$ that governs our quantum particle.

\subsection{A (quantum) holographic dictionary}\label{tenrep}

Inside the screen, spacetime has already emerged. This gives us the energy representation of quantum mechanics---the one we are used to: a time variable with a conserved Noether charge, the energy, and wavefunctions depending on the spacetime coordinates. We have the uncertainty relation
\begin{equation}
\Delta \hat Q\,\Delta \hat P\geq \frac{\hbar}{2}.
\label{ineq}
\end{equation}
Expectation values are computed as functional integrals, with a density function $d_I$ given by
\begin{equation}
d_I=\exp\left(\frac{{\rm i}}{\hbar}I\right).
\label{sabemos}
\end{equation}
Above, $I=\int{\rm d}t L$ is the action integral satisfying the Hamilton--Jacobi equation.

We can now posit the quantum--mechanical analogues of eqns.  (\ref{sabinebiene}), (\ref{tivi}) and (\ref{defitermonova}). In the energy representation these analogues read, respectively,
\begin{equation}
\frac{1}{k_B}\hat S(x)=\frac{1}{4\hbar c L_P}A(U(x)){\vert\hat U(x)\vert},
\label{lieber}
\end{equation}
\begin{equation}
2\pi k_B\hat T(x)=L_P\frac{{\rm d}\hat U}{{\rm d}n},
\label{tivix}
\end{equation}
\begin{equation}
\frac{k_B}{2}\int_{\cal S}{\rm d}^2a\, \hat T=\frac{A({\cal S})}{4\pi}\hat U ({\cal S}).
\label{deficuanto}
\end{equation}

\subsection{Emergence of the holographic correspondence}\label{thentrep}

It is well known, in the theory of thermodynamical fluctuations, that the probability density function $d_S$ required to compute expectation values of thermodynamical quantities is given by the exponential of the entropy \cite{CALLEN}:
\begin{equation}
d_S=\exp\left(\frac{S}{k_B}\right).
\label{scheiss}
\end{equation}
Comparing (\ref{scheiss}) with (\ref{sabemos}) we arrive at the holographic correspondence (\ref{korrespondenzia})
\begin{equation}
\frac{{\rm i}I}{\hbar}\leftrightarrow\frac{S}{k_B}
\label{korres}
\end{equation}
between the energy representation and the entropy representation.

We would like to point out that an analogous correspondence has been given in \cite{ISIDRO}, the only difference being a factor of 2 in the denominator on the right--hand side. This factor of 2 is easy to account for. In \cite{ISIDRO}, one compares the semiclassical wavefunction in the energy representation, given by $\psi=\exp\left({\rm i}I/\hbar\right)$, with the square root of the probability density function $d_S$ in the entropy representation, given by $\sqrt{d_S}=\exp\left(S/2k_B\right)$. Instead, here we are equating the probability densities $d_I$ and $d_S$ rather than the wavefunctions. See refs. \cite{BANERJEE1, BANERJEE2} for specific instances of the correspondence (\ref{korres}).

\subsection{Emergence of the wavefunction}\label{qsvshs}

The equation $U(x^1, x^2, x^3)=U_0$, where $U_0$ is a constant, defines an equipotential surface in $\mathbb{R}^3$. As $U_0$ runs over all its possible values, we obtain a foliation of  $\mathbb{R}^3$ by equipotential surfaces. Following \cite{VERLINDE}, we will identify equipotential surfaces with holographic screens. Hence forces will arise as entropy gradients.

Assume that $\psi$ is nonvanishing at a certain point in space. Consider an infinitesimal cylinder around this point, with height $L_P$ and base area equal to the area element ${\rm d}^2a$. Motivated by the proportionality between area and entropy, already mentioned, we postulate that there is an infinitesimal entropy flow ${\rm d}S$ {\it from the particle to the area element}\/ ${\rm d}^2a$:
\begin{equation}
{\rm d}S=C \,2\pi k_BL_P \vert\psi\vert^2{\rm d}^2a.
\label{helement}
\end{equation}
Here $C$ is a dimensionless numerical constant, to be determined presently. A closed surface $\Sigma$ receives an entropy flux $S(\Sigma)$:
\begin{equation}
S(\Sigma)=C(\Sigma)2\pi k_BL_P \int_{{\Sigma}}{\rm d}^2a\,\vert\psi\vert^2.
\label{hassociat}
\end{equation}
The constant $C(\Sigma)$ will in general depend on the particular surface chosen; the latter may, but need not, be a holographic screen.  The key notion here is that the integral of the scalar field $\vert\psi\vert^2$ over any surface carries an entropy flow associated. When the surface $\Sigma$ actually coincides with a holographic screen ${\cal S}$, and when the latter is not a nodal surface of $\psi$, the constant $C({\cal S})$ may be determined by the requirement that the entropy flux from the particle to the screen equal the quantum of entropy (\ref{entropia}). Thus
\begin{equation}
\frac{1}{C({\cal S})}=L_P \int_{{\cal S}}{\rm d}^2a\,\vert\psi\vert^2.
\label{cequattro}
\end{equation}
Let us now read eqn. (\ref{cequattro}) in reverse, under the assumption that one knows the proportionality constants $C({\cal S}_j)$ for a given foliation $\mathbb{R}^3=\cup_{j\in {\cal J}}{\cal S}_j$. This amounts to a knowledge of  the integrands, {\it i.e.}, of the probability density $\vert\psi\vert^2$ within the surface integral (\ref{cequattro}) on each and every ${\cal S}_j$. From these tomographic sections of all probability densities {\it there emerges the complete wavefunction $\psi$ on all of}\/ $\mathbb{R}^3$, at least up to a (possibly point--dependent) phase ${\rm e}^{{\rm i}\alpha}$.

Thus the integrand of (\ref{cequattro}) gives the surface density of entropy flow into the holographic screen ${\cal S}_j$, and the wavefunction $\psi$ becomes (proportional to) the square root of this flow. The collection of all these tomographic sections of $\psi$ along all possible screens amounts to a knowledge of the complete wavefunction. Hence {\it a knowledge of the different surface densities of entropy flux across all possible screens is equivalent to a knowledge of the quantum--mechanical wavefunction} $\psi$. This is how the quantum--mechanical wavefunction $\psi$ emerges from the holographic screens.

\section{The Unruh equation of state}\label{tfeteosaeq}

In this section we will rewrite  the dictionary entries (\ref{lieber}), (\ref{tivix}) and (\ref{deficuanto}), postulated to hold in the energy representation of quantum mechanics, in the entropy representation. For this purpose we first need to solve the eigenvalue equation $\hat S\phi=S\phi$ on the screen, so the latter will be kept fixed. That is, we will not consider a variable surface ${\cal S}_j$ of the foliation, but  rather a specific surface corresponding to a fixed value of the index $j$. Observe also a difference in notation: $\phi$ instead of $\psi$. This is to stress the fact that, by (\ref{lieber}), entropy eigenstates $\phi$ cannot be eigenstates of the complete Hamiltonian $\hat H$, but only of the potential energy $\hat U$. Once we have solved the eigenvalue equation
\begin{equation}
\hat U\phi=U\phi,
\label{usombrero}
\end{equation}
then the same $\phi$ diagonalise $\hat S$:
\begin{equation}
\hat S\phi=S\phi, \qquad S=\frac{k_B}{4\hbar c L_P}A({\cal S}){\vert U({\cal S})\vert}.
\label{eigenentropia}
\end{equation}
Thermodynamical quantities will now arise as expectation values of operators in the entropic eigenstates $\phi({\cal S})$.

We first deal with (\ref{lieber}). Clearly its reexpression in the entropy representation will be the thermodynamical fundamental equation $S=S(A)$, since the extensive parameter corresponding to the holographic screen is the area $A$. Then we have
\begin{equation}
\langle \hat S\rangle=\frac{k_B}{4\hbar c L_P}A({\cal S}){\vert U({\cal S})\vert}.
\label{biacca}
\end{equation}
Availing ourselves of the freedom to pick the origin of potentials at will, let us set $\vert U({\cal S})\vert=\hbar c/L_P$. Thus
\begin{equation}
\langle \hat S\rangle=\frac{k_B}{4L_P^2}A,
\label{bhache}
\end{equation}
which is the celebrated Bekenstein--Hawking law. It arises as a thermodynamical fundamental equation in the entropy representation.

Our holographic screen is treated thermodynamically as a stretched membrane, so the generalised force conjugate to the extensive parameter $A$ is the surface tension $\sigma$. Then the equation of state corresponding to (\ref{bhache}) is
\begin{equation}
\sigma=\frac{k_B\langle \hat T\rangle}{4L_P^2}.
\label{estado}
\end{equation}
Rewrite the above as $2\pi k_B\langle \hat T\rangle=8\pi L_P^2\sigma$ and recall that $\sigma$ is the normal component of force per unit length on the screen. Since force is proportional to acceleration, the above equation of state turns out to be equivalent to the Unruh law.

{}Finally we turn to the first law of thermodynamics and the equipartition theorem. Taking the expectation value, in the entropic eigenstates $\phi$, of the operator equation (\ref{deficuanto}), produces the thermodynamical expression for the equipartition theorem:
\begin{equation}
\frac{k_B}{2}\int_{\cal S}{\rm d}^2a\, \langle\hat T\rangle=\frac{A({\cal S})}{4\pi}\langle\hat U ({\cal S})\rangle.
\label{equinova}
\end{equation}

\section{The second law of thermodynamics, revisited}

The second law of thermodynamics,
\begin{equation}
\Delta S\geq 0,
\label{segunda}
\end{equation}
has been related to the Heisenberg uncertainty principle in ref. \cite{OLAH}. In ref. \cite{DURHAM} it has been argued that the second law of thermodynamics has a quantum--mechanical reexpression in the Bell inequalities. In ref. \cite{ISIDRO} we have established a link between (\ref{segunda}) and the Hilbert space of entropic quantum mechanics. Here we would like to propose yet another quantum--mechanical interpretation of the second law, one that combines the uncertainty principle with the notion of emergence.

{}From eqn. (\ref{entropia}) one derives the obvious inequality
\begin{equation}
\Delta S\geq \pi k_B
\label{triqui}
\end{equation}
which looks like some refinement of the second law (\ref{segunda})---the latter would be recovered in the semiclassical limit $k_B\to 0$.
Therefore let us, for the sake of the argument, consider eqn. (\ref{triqui}) as a more precise statement of the second law than (\ref{segunda}). As such (\ref{triqui}) is reminiscent the uncertainty principle (\ref{ineq}) of quantum mechanics. However the left--hand side of (\ref{triqui}) contains just one uncertainty, instead of a product of two uncertainties as usual. This reflects the fact that the variable on the left, $S$, is {\it selfconjugate}\/---its dimension equals that of the quantum $k_B$ on the right--hand side\footnote{Compare this situation with $(q,p)$ and $(H, t)$, which are conjugate pairs: the product of the two components of each pair has the dimension of $\hbar$. Angular momentum $L$ is selfconjugate, in the sense that it carries the dimension of $\hbar$, but one writes the corresponding uncertainty principle as $\Delta L \Delta \varphi\geq \hbar/2$, where the dimensionless variable $\varphi$ is an angle.}. We can include a dimensionless formal parameter $\tau$ in the left--hand side that will make (\ref{triqui}) resemble the uncertainty principle in its standard form. This can be done as follows.

Let $N$ denote the total number of bits on ${\cal S}$.  Whenever a quantum particle hits the screen we have $\Delta N=1$, and the ratio $\Delta N/N$ will be small if $N$ is large enough. In this limit we can treat $N$ as a continuous variable, that we redenote by $\tau$ in order to interpret it as a continuous, dimensionless parameter:
\begin{equation}
\tau:=N, \qquad {\rm when}\qquad \frac{\Delta N}{N}<<1.
\label{thau}
\end{equation}
This is the limit $N\to\infty$ referred to in (\ref{numm}). Compatibility with all the above requires this limit to correspond to $k_B\to 0$ or, equivalently, to $\hbar\to 0$.
In other words, the large area limit for a holographic screen corresponds to the semiclassical approximation in quantum mechanics.

We have $\Delta\tau\geq 1$, the inequality allowing for the possibility of more than just one particle hitting ${\cal S}$. Thus multiplying the two inequalities $\Delta \tau\geq 1$ and $\Delta S\geq \pi k_B$ together we arrive at the following uncertainty principle {\it on the holographic screen}\/:
\begin{equation}
\Delta S \Delta \tau\geq \pi k_B.
\label{ungewiss}
\end{equation}
The fact that $k_B$, though small, is nonvanishing, leads to the impossibility of having strictly reversible processes; reversibility is possible only in the limiting case of a vanishing value for the quantum $k_B$. We conclude that quantisation appears as dissipative mechanism. The notion that information loss leads to a quantum behaviour lies at the heart of the notion of emergence \cite{VITIELLO1, VITIELLO2, ELZE1, ELZE2, GROESSING0, GROESSING1, GROESSING2, GROESSING5, THOOFT1, THOOFT2, VITIELLO3}.

We have derived the uncertainty principle (\ref{ungewiss}) starting from the second law of thermodynamics (\ref{segunda}). Let us now prove that the reverse path is also possible: from the uncertainty principle to the second law of thermodynamics. We start from (\ref{ineq}) in the bulk rewritten as $\Delta I/\hbar\geq 1$, where $I=\int p{\rm d}q$ is the action. On the boundary, the correspondence (\ref{korrespondenzia}) allows to reexpress the above inequality as in (\ref{triqui}). Along the way we have dropped irrelevant numerical factors.

Altogether, we have an equivalence between the uncertainty principle of quantum mechanics (either in the bulk (\ref{ineq}) or on the boundary (\ref{ungewiss})), and a refined version of the second law of thermodynamics, one that includes a small but nonvanishing value of the corresponding quantum ($\hbar$ or $k_B$) on the right--hand side. This is in agreement with the results of \cite{OLAH}---now with the added bonus that our equivalence has the properties of emergence and holography.

\section{Discussion}

The entropy representation of quantum mechanics, as presented here, is a holographic projection of the energy representation of the same theory, as defined on spacetime.
Our central claim, summarised by eqn. (\ref{korrespondenzia}), expresses this holographic property.

There is, however, one additional property of quantum mechanics that is deeply encoded in eqn. (\ref{korrespondenzia}); as such it is not immediately recognised. Namely, quantum mechanics is an emergent phenomenon {\it also}\/ because quantum mechanics is defined on spacetime, and spacetime itself is an emergent phenomenon. Let us analyse this latter point in more detail.

Any model of emergent gravity must ultimately account for the laws governing the motion of material bodies. Thus, {\it e.g.}, the proposal made in \cite{VERLINDE} allows for a (somewhat heuristic) derivation of Newton's law of motion, $F=ma$, and of the relativistic generalisations thereof, as emergent, thermodynamical laws. Moreover, the intriguing presence of Planck's constant $\hbar$ \cite{CHEN} in the purely classical setup of ref. \cite{VERLINDE} makes one suspect that quantum mechanics also has a role to play in that setup. On the other hand, it is well known that Newton's law $F=ma$ can be recovered in the semiclassical limit of quantum mechanics, as being satisfied by the expectation values of certain operators (Ehrenfest's theorem). Last but not least, thermodynamics is the paradigm of emergent phenomena.

All these different pieces of evidence point toward one and the same conclusion---{\it viz.}, that if classical mechanics follows from the emergence property of spacetime, then the same should be true of quantum mechanics. Here and in ref. \cite{ISIDRO} we have exploited this point of view. We would like to stress that this conclusion is ultimately independent of the precise mechanism whereby spacetime emerges. Thus, although the holographic dictionary presented in previous sections hinges crucially on the emergence mechanism being precisely that of ref. \cite{VERLINDE}, the holographic correspondence (\ref{korrespondenzia}) is independent of that mechanism. As such, the holographic correspondence (\ref{korrespondenzia}) should hold just as well in any other specific model for the emergence of spacetime (say, loop quantum gravity or any alternative thereto such as \cite{FINSTER, MAKELA}).

\vskip.5cm
\noindent
{\bf Acknowledgements}  J.M.I. heartily thanks the organisers of the {\it 5th International Heinz von Foerster Congress: Emergent Quantum Mechanics}\/, Vienna, Nov. 11-13,  2011,  for the opportunity to present this talk, and for providing a congenial environment for scientific discussions. This work has been supported by Universidad Polit\'ecnica de Valencia under grant PAID-06-09.\\
{\it Handle stets so, da{\ss}  weitere M\"oglichkeiten entstehen.\\
---Heinz von Foerster.}

\end{document}